\documentclass[iop,apj,a4paper]{emulateapj}
\usepackage{apjfonts,amsmath,amssymb,epsf}
\submitted{Received 2012 December 21; accepted 2013 February 13; published 2013 ?? ??}

\begin{document}

\title{A  Gravitational Lens Model for the Ly$\alpha$ Emitter, LAE
221724$+$001716 at $z$ = 3.1 in the SSA 22 Field\altaffilmark{1,2}}

\author{Y. Nakahiro\altaffilmark{3},
Y. Taniguchi\altaffilmark{4},
A. K. Inoue\altaffilmark{5},
Y. Shioya\altaffilmark{4},
M. Kajisawa\altaffilmark{3,4},
M. A. R. Kobayashi\altaffilmark{4},
I. Iwata\altaffilmark{6},
Y. Matsuda\altaffilmark{7},
T. Hayashino\altaffilmark{8},
A. R. Tanaka\altaffilmark{3}, and
K. Hamada\altaffilmark{3}}

\altaffiltext{1}{Based on data collected at the Subaru Telescope,
which is operated by the National Astronomical Observatory of Japan.}
\altaffiltext{2}{This paper is dedicated to Mr. Masahiro Nakahiro (the
father of the first author), who deceased on the 21st December 2012.}
\altaffiltext{3}{Graduate School of Science and Engineering, Ehime
University, Bunkyo-cho 2-5, Matsuyama, Ehime 790-8577, Japan}
\altaffiltext{4}{Research Center for Space and Cosmic Evolution, Ehime
University, Bunkyo-cho 2-5, Matsuyama, Ehime 790-8577, Japan}
\altaffiltext{5}{College of General Education, Osaka Sangyo
University, 3-1-1, Nakagaito, Daito, Osaka 574-8530, Japan}
\altaffiltext{6}{Subaru Telescope, National Astronomical Observatory
of Japan, 650 North A'ohoku Place, Hilo, HI 96720, USA}
\altaffiltext{7}{ALMA Office, National Astronomical Observatory of
Japan, 2-21-1 Osawa, Mitaka, Tokyo 181-8588, Japan}
\altaffiltext{8}{Research Center for Neutrino Science, Graduate School
of Science, Tohoku University, Aramaki, Aoba-ku, Sendai 980-8578,
Japan}

\email{kobayashi@cosmos.phys.sci.ehime-u.ac.jp}

\begin{abstract}

During the course of our Lyman continuum imaging survey, we found that
the spectroscopically confirmed Ly$\alpha$ emitter LAE 221724$+$001716
at $z = 3.10$ in the SSA 22 field shows strong Lyman continuum
emission ($\lambda_{\rm rest}\sim 900$~{\AA}) that escapes from this
galaxy.  However, another recent spectroscopic survey revealed that
the supposed Lyman continuum emission could arise from a foreground
galaxy at $z = 1.76$ if the emission line newly detected from the
galaxy at $\lambda_{\rm obs} \approx 3360$~{\AA} is Ly$\alpha$.  If
this is the case, as the angular separation between these two galaxies
is very small ($\approx 0.6^{\prime\prime}$), LAE 221724$+$001716 at
$z = 3.10$ could be amplified by the gravitational lensing caused by
this intervening galaxy.  Here we present a possible gravitational
lens model for the system of LAE 221724$+$001716.  First, we estimate
the stellar mass of the intervening galaxy as $M_\star \sim 3.5 \times
10^9~M_\odot$ from its UV luminosity and $\sim 3.0\times
10^7$--$2.4\times 10^9~M_\odot$ through the SED fitting.  Then, we
find that the gravitational magnification factor ranges from 1.01 to
1.16 using the so-called singular isothermal sphere model for strong
lensing.  While LAE 221724$+$001716 is the first system of an LAE-LAE
lensing reported so far, the estimated magnification factor is not so
significant because the stellar mass of the intervening galaxy is
small.

\end{abstract}


\section{INTRODUCTION}

Toward resolving the cosmic reionization sources, observations
directly detecting the Lyman continuum ($\lambda_{\rm rest} <
912$~{\AA}, hereafter LyC) from galaxies at $z>3$ have proceeded in
the last decade (Steidel et al. 2001; Giallongo et al. 2002;
Fern{\'a}ndez-Soto et al. 2003; Inoue et al. 2005; Shapley et
al. 2006; Iwata et al. 2009; Vanzella et al. 2010a; Boutsia et
al. 2011; Nestor et al. 2011, 2012). Iwata et al. (2009) and Nestor et
al. (2011, 2012) reported the largest number of individual detections
of the LyC from galaxies in the SSA22 field where a massive
proto-cluster of galaxies at $z=3.10$ was found (Steidel et al. 1998;
Hayashino et al. 2004).  Extremely strong LyC detected from some
sample galaxies suggested that they contain a primordial stellar
population like metal-free stars with mass fraction of 1--10\% (Inoue
et al. 2011). However, Vanzella et al. (2010b) raised a possibility
that the supposed LyC detected from the galaxies come from a
foreground galaxy closely aligned along the line-of-sight toward the
background $z\sim3$ galaxies.  The LyC observers discussed this
possibility in their papers and concluded that it was unlikely that
all the detections were foreground contamination, while a part of them
were so (Iwata et al. 2009; Nestor et al. 2011, 2012).  Indeed, Nestor
et al. (2012) found an example of such a contamination: a foreground
galaxy very closely ($ < 1^{\prime \prime}$) located in the front of a
Ly$\alpha$ emitter (LAE) with the possible LyC detection reported
(Iwata et al. 2009; Inoue et al. 2011; Nestor et al. 2011).
Hereafter, we call this galaxy LAE 221724$+$001716.

Because of the small angular separation between the foreground and
background galaxies, LAE 221724$+$001716 may be a strong gravitational
lensing system.  The gravitational lensing is now recognized as a
powerful tool for observational cosmology and galaxy formation and
evolution studies (e.g., Kneib \& Natarajan 2011; see also Coe et
al. 2013 and references therein).  The strong lensing enables us to
search faint high-redshift galaxies behind galaxy clusters and
investigate physical properties of galaxies with higher
signal-to-noise ratio by brightening and magnifying effect.  If LAE
221724$+$001716 is a strongly magnified $z \sim 3$ LAE, the intrinsic
luminosity should be very faint, whereas its apparent luminosity is
brightest among general LAEs.  In this case, this object provides us
with a chance to investigate the nature of a faint LAE in detail.
Therefore, we try to construct a simple lensing model and estimate the
magnification factor of this possible lensing system in this paper.

Throughout this paper, magnitudes are given in the AB system.  We
adopt a flat universe with $\Omega_{\rm M} = 0.3$, $\Omega_\Lambda =
0.7$, and $H_0 = 70~ {\rm km~ s^{-1}~ Mpc^{-1}}$.

\section{LAE 221724$+$001716}

\begin{figure*}
\epsscale{.90}
\plotone{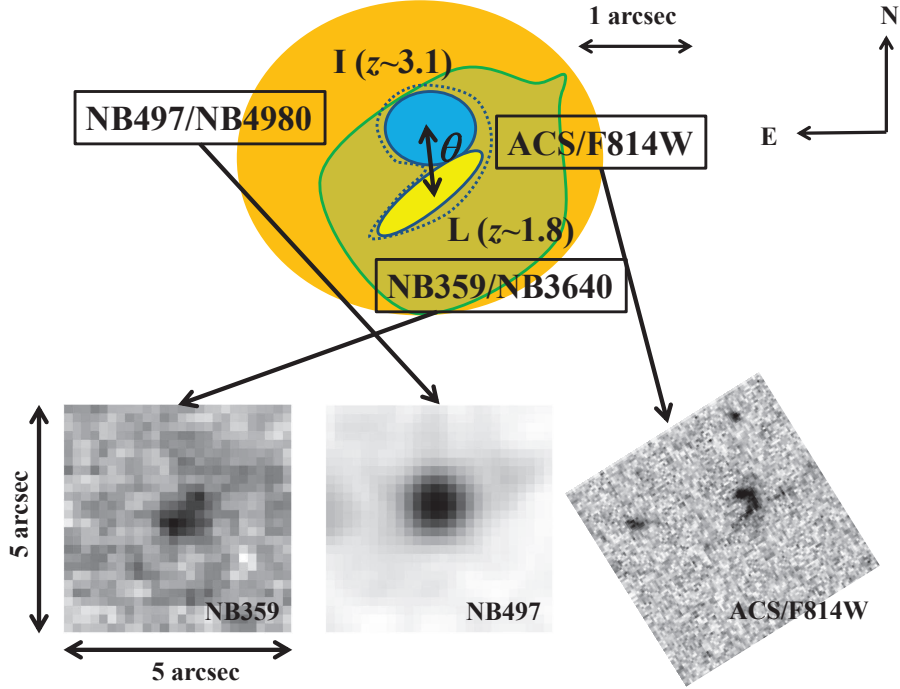}
\caption{
Schematic view of the relative positions of LAE 221724$+$001716 and
the lensing galaxy.  The light blue ellipse labeled ``I'' shows the
image of LAE 221724$+$001716 at $z=3.10$ and the yellow ellipse
labeled ``L'' shows the foreground lensing galaxy at $z=1.76$.  The
orange area shows the $NB497/NB4980$ emission that is dominated by the
Ly$\alpha$ emission from LAE 221724$+$001716 and the green area shows
the $NB359/NB3640$ emission that is dominated by the rest-frame UV
stellar continuum emission from the foreground galaxy.  The angular
separation $\theta$ is defined as the angle between the
luminosity-weighted central positions of the $NB3640$ and $NB4980$
images ($= 0.6^{\prime\prime}$; Nestor et al. 2011) or $NB359$ and
$NB497$ images ($= 0.38^{\prime\prime}$; Inoue et al. 2011).  For
reference, the thumbnail images for $NB359$, $NB497$, and $ACS/F814W$
are shown in the bottom column.  North is up and east is left.  The
field of view is $5^{\prime \prime}\times 5^{\prime \prime}$.
\label{fig:fonfiguration}}
\end{figure*}
LAE 221724$+$001716 is an LAE found in the SSA22 field.  Its strong
Ly$\alpha$ emission line at $z =3.10$ was confirmed by spectroscopy
(Matsuda et al. 2006; Iwata et al. 2009; Inoue et al. 2011; Nestor et
al. 2012).  Its $NB497$ (central wavelength $\lambda_{\rm c} =
4977$~{\AA}, full width at the half maximum $\Delta \lambda =
77$~{\AA}) magnitude, $NB497 = 23.78$, is one of the brightest among
the LAE candidates in the SSA22 field (Yamada et al. 2012).  This
object is also classified as a Ly$\alpha$ blob (LAB; No. 35) in
Matsuda et al. (2004) because of its spatially extended Ly$\alpha$
emission, but this spatial extent is caused by the connection of the
emission line from an eastern close companion object (the separation
between them is $\sim 3^{\prime \prime}$).  In fact, around this LAE,
there are four objects within 4$^{\prime\prime}$: eastern, north,
north-western, and western objects as seen in the HST/ACS I band image
(see Iwata et al. 2009; Nestor et al. 2011, 2012).

The shape of the central object in the HST/ACS image appears to be
peculiar: {\it ``maga-tama''} (ancient Japanese amulet made of stone)
like shape.  If we look closely at this object, it seems to be
composed of two parts as northern {\it ``head''} and southern {\it
``tail''}.  Iwata et al.  (2009) detected this object in their $NB359$
($\lambda_{\rm c} = 3590$~{\AA}, $\Delta \lambda = 150$~{\AA}) image
tracing the LyC at $z=3.10$ and Inoue et al. (2011) named this object
{\it ``I11-a''}.  Nestor et al. (2011) confirmed this detection by
their $NB3640$ ($\lambda_{\rm c}=3630$~{\AA}, $\Delta \lambda =
100$~{\AA}) imaging and named it as {\it ``LAE003''}.  There is a
slight spatial offset between the $NB359/NB3640$ (supposed LyC from
LAE 221724$+$001716 at $z=3.10$) and $NB497/NB4980$ ($\lambda_{\rm c}
= 4985$~{\AA}, $\Delta \lambda = 80$~{\AA}) (Ly$\alpha$) positions:
0.4--0.6$^{\prime\prime}$.  From this small offset, Inoue et
al. (2011) suggested a very low possibility that $NB359/NB3640$ flux
comes from a foreground object.  However, Nestor et al. (2012) found
an emission line around 3360~{\AA} in the new deep spectrum of this
object, strongly suggesting the presence of a foreground galaxy very
closely aligned toward the $z=3.10$ LAE.  If this emission line is
Ly$\alpha$, the foreground galaxy lies at $z=1.76$.  Nestor et
al. (2012) also noted that the southern tail is the foreground galaxy
because of a small spatial offset of the 3360~{\AA} line and
4980~{\AA} line (Ly$\alpha$ at $z=3.10$).  The geometrical
configurations of this system are shown in Figure~1.

The angular separation between LAE 221724$+$001716 detected in
$NB497/NB4980$ bands and the foreground galaxy detected in the
$NB359/NB3640$ bands is estimated as $0.38^{\prime\prime}$ (Inoue et
al.  2011) and $0.6^{\prime\prime}$ (Nestor et al. 2011).  While these
two values are basically consistent within their observed errors,
$\approx 0.2^{\prime\prime}$, we adopt the observed offset reported by
Nestor et al. (2011), $0.6^{\prime\prime}$, as the fiducial angular
separation between the two objects.  This is because the
luminosity-weighted central position of an extended source depends on
the depth of the image and the $NB3640$ and $NB4980$ images of Nestor
et al. (2011) are significantly deeper ($\sim 0.57$--1.2~mag) than the
$NB359$ and $NB497$ images of Inoue et al. (2011).  We also present
the results in the case of the angular separation reported by Inoue et
al. (2011) for reference in Section 4.


\begin{table*}
 \begin{center}
  \caption{Photometric data of LAE 221724$+$001716.  All flux
  densities are measured within 1.2$^{\prime\prime}$ diameter aperture
  at the position in $R$. The aperture correction factor is 2.43 for
  the estimation of total flux densities.
  }
  \label{tb:parameters}
  \begin{tabular}{llccl}
   \hline\hline
Band & Telescope/Instrument & $f_\nu$ & $m_{\rm AB}$ & References \\
     & & ($\rm \mu Jy$) &     \\
\hline
$NB359$ & Subaru/Suprime-Cam & $0.1526 \pm 0.0130$ & 25.94 & Iwata et al. (2009)\\
$u^*$ & CFHT/MegaCam       & $0.1832 \pm 0.0061$ & 25.74 & Kousai (2011) \\
$B$   & Subaru/Suprime-Cam & $0.2163 \pm 0.0045$ & 25.56 & Hayashino et al. (2004)\\
$NB497$ & Subaru/Suprime-Cam & $0.5123 \pm 0.0062$ & 24.63 & Hayashino et al. (2004)\\
$V$   & Subaru/Suprime-Cam & $0.2378 \pm 0.0044$ & 25.46 & Hayashino et al. (2004)\\
$R$   & Subaru/Suprime-Cam & $0.2271 \pm 0.0043$ & 25.51 & Hayashino et al. (2004)\\
$i^\prime$ & Subaru/Suprime-Cam & $0.2420 \pm 0.0049$ & 25.44 & Hayashino et al. (2004)\\
$z^\prime$ & Subaru/Suprime-Cam & $0.2452 \pm 0.0072$ & 25.43 & Hayashino et al. (2004)\\
$J$   & Subaru/MOIRCS & $< 0.4365$ & $> 24.80$ & Uchimoto et al. (2008)\\
$H$   & Subaru/MOIRCS & $< 0.3192$ & $> 25.14$ & Uchimoto et al. (2008)\\
$K_{\rm s}$ & Subaru/MOIRCS & $< 0.5395$ & $> 24.57$ & Uchimoto et al. (2008)\\
\hline
  \end{tabular}
 \end{center}
\end{table*}
Before moving the next section, we here summarize the photometry of
this object.  We measure its flux densities in 11 bands from optical
to near-infrared.  The details of the observations with the Subaru
telescope and the data reductions are described in Hayashino et
al. (2004) for $B$, $V$, $R$, $i'$, $z'$, and $NB497$, Uchimoto et
al. (2008) for $J$, $H$, and $K$, and Iwata et al. (2009) for $NB359$.
The $u^*$ image were taken with the CFHT/MegaCam and described in
Kousai (2011).  The spatial positions of this object in $NB359$ and
$NB497$ are slightly ($<0.5^{\prime\prime}$) offset from it in $R$ as
mentioned above.  Nevertheless, we simply applied aperture photometry
at the $R$ position with $1.2^{\prime\prime}$ diameter.  Table~1 is
the result.  As an estimate of the total flux density, we adopt
MAG\_AUTO by SExtractor (Bertin \& Arnouts 1996).  Then, we find the
aperture correction factor of 2.43 (i.e., $-0.96$~mag) in $R$.  We
apply it even for other bands when estimating the total flux densities
in the following.

\section{GRAVITATIONAL LENS MODEL FOR LAE 221724$+$001716}

We construct a gravitational lens model for LAE 221724$+$001716.  We
use the so-called Singular Isothermal Sphere (SIS) lens model (e.g.,
Binney \& Merrified 1998) for LAE 221724$+$001716.  This lens model is
appropriate for describing the actual density distribution of the dark
matter halo of a galaxy.  Moreover, this SIS lens model has only one
parameter, that is, the velocity dispersion for the foreground galaxy,
$\sigma$.  Thus, it is also convenient to estimate the gravitational
magnification factor that is given below:
\begin{equation}
M_+ = \frac{\theta}{\theta - \theta_{\rm E}},
\end{equation}
where $\theta$ is the angular separation between the source and the
lensed image, $\theta = 0.6^{\prime\prime}$ (Nestor et al. 2011).  The
Einstein angle, $\theta_{\rm E}$, is defined as
\begin{equation}
\theta_{\rm E}= \frac{4\pi\sigma^2} {c^2} \frac{D_{\rm LS}} {D_{\rm OS}},
\end{equation}
where $\sigma$ is the velocity dispersion of dark matter halo hosting
the lensing galaxy, $D_{\rm LS}$ is the angular diameter distance
between the foreground galaxy and LAE 221724$+$001716, and $D_{\rm
OS}$ is that between the observer and LAE 221724$+$001716; see
Figure~2.
\begin{figure}
\epsscale{1.2}
\plotone{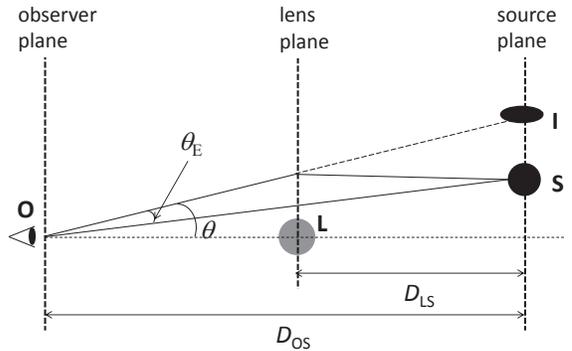}
\caption{
Geometry of a gravitational lens system.  Here $D_{\rm LS}$ and
$D_{\rm OS}$ are the angular diameter distances between the lensing
galaxy at $z=1.76$ (labeled ``L'') and LAE 221724$+$001716 at $z=3.10$
(labeled ``S'') and that between the observer (labeled ``O'') and LAE
221724$+$001716, respectively.  The angular separation between the
lensed image (labeled ``I'') and the lensing galaxy is represented by
$\theta$.
\label{fig:lensmodel}}
\end{figure}


In order to estimate $\sigma$, we use the stellar-mass Tully-Fisher
relation (Pizagno et al. 2005; Swinbank et al. 2012),
\begin{equation}
V_{2.2} = 155.6 \left( \frac{\Delta TF}{2.32} \frac{M_\star}{10^{10}~
		 M_\odot} \right)^{1/3.05}~ {\rm km~s^{-1}},
\label{eq-Sigma+Mstar}
\end{equation}
where $V_{2.2}$ is the rotation velocity at 2.2 times the disk scale
length and $\Delta TF$ is the evolution factor of the zero-point of
the Tully-Fisher relation from $z \sim 0$ to $z \sim 2$ defined by
$M_{\star}(z \sim 0)/M_{\star}(z \sim 2)$ (see Fig. 6 of Swinbank et
al. 2012 in detail).  We also note that this relation is valid for the
Chabrier initial mass function (IMF) with 0.1--100~$M_\odot$.

Now we estimate the stellar mass of the lensing galaxy ($M_{\star,
{\rm L}}$), which is used as an input to eq.~(\ref{eq-Sigma+Mstar}),
using the following two independent methods: (A) the UV luminosity -
stellar mass relation (Sawicki 2012), and (B) the spectral energy
distribution (SED) fitting method.  First, we use the Sawicki
relation,
\begin{equation}
\log_{10}{\left(\frac{M_\star}{M_\odot}\right)} = 0.68 - 0.46M_{\rm
 UV},
\label{eq-Sawicki}
\end{equation}
where $M_{\rm UV}$ is the absolute UV magnitude at the rest-frame
wavelength of 1700~{\AA}.  This relation is empirically obtained from
the UV-selected galaxies adopting the Salpeter IMF with
0.1--100~$M_\odot$.  The stellar mass obtained with the Salpeter IMF
is larger than that with the Chabrier IMF by a factor of 1.8.  The
observed total AB magnitude of $NB3640 = 24.74$ by Nestor et
al. (2011) gives $M_{\rm UV} = -19.78$.  While this corresponds to the
magnitude at $\lambda_{\rm rest} = 1315$~{\AA}, it can be regarded as
the magnitude at $\lambda_{\rm rest} = 1700$~{\AA} under the
assumption that the lensing galaxy has a flat continuum.  This
assumption is reasonable since the lensing galaxy is an LAE and such a
flat continuum is expected for the galaxies with young age and a small
amount of dust like LAE.  From the Sawicki relation, we obtain
$M_{\star, {\rm L}} = 3.5 \times 10^9 M_\odot$ for the Chabrier IMF.
We adopt this value in later discussion.

Next, we use the SED fitting method.  Here we use Bruzual \& Charlot
(2003) model together with the following parameters: (1) Chabrier IMF
with $0.1$--100~$M_\odot$, (2) the metallicity $Z = 0.004$ ($ =
0.2~Z_\odot$), (3) constant star formation history and (4) the
extinction curve by Calzetti et al. (2000) with $E(B-V) = 0$--0.2 with
an interval of 0.05.  In our analysis, we use the total flux densities
in optical and near infrared wavelengths described in Section 2.  The
total flux densities are a combination of those from LAE
221724$+$001716 at $z=3.10$ and the foreground galaxy at $z=1.76$.
However, the total flux density of $NB359$ is considered to be
dominated by the flux from the foreground galaxy if the escape
fraction of the LyC photons from LAE 221724$+$001716 is not
significant.  Hence, in the course of the SED fitting to determine a
robust upper limit and possible lower limit of $M_{\star, {\rm L}}$,
we use the total flux density of $NB359$ with its photometric error
and regard those of the other bands as upper limits because they are
contaminated by the flux from LAE 221724$+$001716.  The resultant
upper and lower limits of $M_{\star, {\rm L}}$ are $2.4\times
10^9~M_\odot$ and $3.0\times 10^7~M_\odot$, respectively.  The upper
limit of $M_{\star, {\rm L}}$ is smaller than that estimated from the
Sawicki relation above (i.e., $M_{\star, {\rm L}} = 3.5\times 10^9\
M_\odot$) by a factor of $\sim 1.5$.  However, this difference is not
surprising because the age of the lensing galaxy is young, $\sim
100$~Myr, and Sawicki (2012) has found that the stellar mass of such
young galaxies are typically much smaller than those estimated from
eq.~(\ref{eq-Sawicki}) at a fixed $M_{\rm UV}$ (see Fig.~6 of Sawicki
2012).

In summary, (A) the Sawicki relation gives $\sigma = V_{2.2} = 105\
{\rm km\ s^{-1}}$.  On the other hand, (B) the SED fitting method
gives $\sigma \simeq 22$--$93~{\rm km\ s^{-1}}$ for $M_{\star,{\rm L}}
\sim 3.0\times 10^7$--$2.4\times 10^9~M_\odot$.  Based on these
results, we estimate the gravitational magnification factor, $M_+$.
In Figure~3, we show our results for the separation between the LAE
and the foreground galaxy, $0.6^{\prime\prime}$.  We obtain $M_+
\simeq 1.16$ and $\simeq 1.01$--1.12 in the cases (A) and (B),
respectively.  We note that the estimated magnification factors are
fairly small although the angular separation between the lensing
galaxy and the source galaxy is very small ($\approx 0.6$~arcsec).
This is due to the low mass of the lensing galaxy at $z \sim 1.8$,
$\sim 10^9\ M_\odot$.
\begin{figure}
\epsscale{1.10}
\plotone{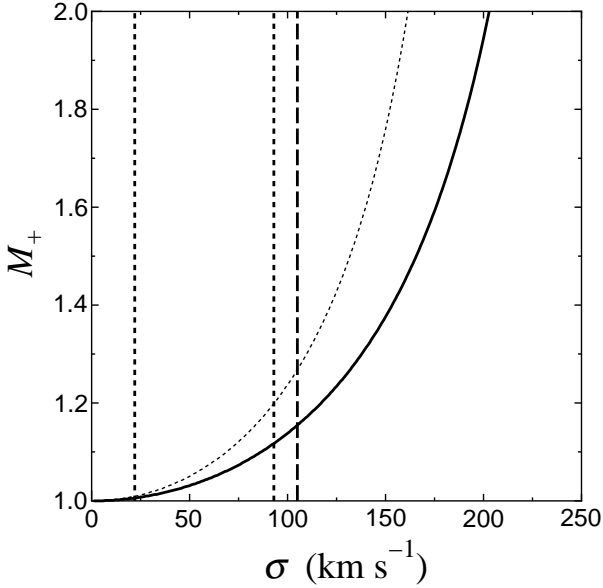}
\caption{
Magnification factor $M_+$ as a function of $\sigma$.  Solid curve and
dotted curves show the case of $\theta = $ $0.6^{\prime\prime}$
(Nestor et al. 2011) and $0.38^{\prime\prime}$ (Inoue et al. 2011),
respectively.  Vertical dotted and dashed lines show $\sigma = $
22--93~${\rm km~s^{-1}}$ and $\sigma = 105~{\rm km~s^{-1}}$ which are
inferred from the SED fitting and the Sawicki relation, respectively.
\label{fig:mplus}}
\end{figure}


\section{DISCUSSION}

In this paper, we have investigated possible models for gravitational
lensing for LAE 221724$+$001716 at $z = 3.10$.  The intervening galaxy
is located at $z = 1.76$ with a very small angular separation of
0.6~arcsec from the line of sight toward LAE 221724$+$001716.  This
small separation suggest that LAE 221724$+$001716 could be amplified
by the gravitational lensing.  If this is the case, LAE
221724$+$001716 is the first system of an LAE-LAE lensing.  However,
our analysis has shown that the magnification factor is only 1.16 at
most.  This result does not change significantly even if we adopt the
smaller angular separation reported by Inoue et al. (2011), $\theta =
0.38^{\prime\prime}$: in this case, $M_+ \simeq 1.27$ and 1.01--1.20
for $\sigma = 105~{\rm km\ s^{-1}}$ and $\sigma = 22$--93~${\rm km\
s^{-1}}$ inferred from the Sawicki relation and the SED fitting,
respectively (see Figure~3).

This small magnification factor is due to that the lensing galaxy at
$z\sim 1.8$ is significantly less massive, i.e., $M_{\star,{\rm L}}
\sim 10^9\ M_\odot$.  It does not change significantly even if we
adopt another SED fitting method with two galaxy components (i.e., the
foreground galaxy at $z=1.76$ and LAE 221724$+$001716 at $z=3.10$) by
using all of the total flux densities; the resultant stellar masses of
the foreground galaxy and LAE 221724$+$001716 are $\sim 7\times
10^8~M_\odot$ and $1\times 10^9~M_\odot$, respectively, in the case of
their SEDs are identical with each other.  However, this stellar mass
is not unusually small among the LAE at $z=3.1$--$6.6$.  On the
contrary, it is similar to that of the LAEs at $z=3.1$ undetected in
$K$-band (Ono et al. 2010).  Hence, the intervening galaxy at $z\sim
1.8$ (and perhaps, the source galaxy at $z\sim 3.1$) is considered to
be a typical LAE.

Regarding the dark halo mass of LAEs, Ouchi et al. (2010) found that
its average inferred from a clustering analysis is roughly $\sim
10^{11\pm 1}~M_\odot$ at $z=2.1$--6.6.  They also reported that the
dark halo masses of LAEs show no significant redshift evolution in
this redshift range beyond the mass-estimate scattering.  It is
possible that the intervening galaxy at $z\sim 1.8$ with
$M_{\star,{\rm L}} \sim 10^9\ M_\odot$ is embedded in such a small
mass halo.  This may imply that the no evolution of halo mass of LAEs
reported by Ouchi et al. (2010) extends to even lower redshift, $z\sim
1.8$.

The narrow-band magnitude of LAE 221724$+$001716 is bright among the
sample of the LAEs at $z\sim 3$ in the SSA22 field given in Yamada et
al. (2012).  Although such property could be attributed to the
gravitational magnification, the small magnification factor obtained
here implies that this LAE is indeed bright.  However, a number of
high-redshift galaxies and quasars could be gravitationally magnified
by the strong lensing either by a foreground galaxy (e.g., Faure et
al. 2008; Muzzin et al. 2012; Inada et al. 2012).  Therefore, it is
worthwhile estimating how often such a lensing event occurs for LAEs
at $z \sim 3$.  Here, we estimate the possibility of an LAE at $z \sim
3$ is gravitationally magnified by a foreground galaxy with $U$-band
magnitude of $\sim 25$.  This is actually the case for LAE
221724$+$001716 studied in this paper since $NB3640 = 24.74$ and
$NB359 = 24.98$ approximately correspond to $U = 25$.  We evaluate the
possibility to find more than one foreground galaxy within 0.6~arcsec
using the $U$-band number count data obtained by VLT/VIMOS (Vanzella
et al. 2010b); note that the central wavelengths of both $NB359$ and
$NB3640$ is similar to that of $U$-band on VLT/VIMOS.  The
possibility, $P_{\rm foreground}$, is calculated by using the
following equation
\begin{equation}
P_{\rm foreground} = \pi r^2 N,
\end{equation}
where $r$ is the radius from the background galaxy and $N$ is the
$U$-band galaxy number count.  Adopting $r=0.6^{\prime\prime}$ and $N
= 57,300\ {\rm deg^{-2}}$ (galaxies with $U=24.5$--25.5), we obtain
$P_{\rm foreground} \simeq 0.005$.  We conclude that the possibility
to find a foreground galaxy around an LAE at $z \sim 3$ within a
radius of 0.6~arcsec is very small; the probability of LAE-LAE lensing
is even smaller if LAE fraction at $z\sim 1.8$ among the entire galaxy
population is small.

Nevertheless, we have found such a case.  A possible reason is that
the $NB359$ detection more preferentially selects foreground galaxies
than blind survey.  If most of $z\sim 3$ galaxies do not emit LyC at
all, they do not detect in $NB359$.  Thus, the $NB359$ detection is
biased toward close-foreground systems (Vanzella et al. 2010b).
However, the number of the $NB359$ detections is significantly larger
than that of the foreground galaxies expected, and thus, some $NB359$
sources are real LyC emitters (Iwata et al. 2009; Inoue et al.  2011;
Nestor et al. 2011, 2012).  In any case, it is interesting that the
LyC survey with $NB359$ imaging is also useful to find
close-foreground systems discussed in this paper.

One interesting property of the gravitational lensing in LAE
221724$+$001716 is that the lensing galaxy is located at $z \sim 1.8$.
Although a number of the gravitational lensing events by a galaxy have
been found to date (e.g., Faure et al. 2008; Muzzin et al. 2012), the
redshifts of the lensing galaxies are preferentially $z < 1$. One
high-redshift lensing galaxy at $z \sim$ 1.5 -- 2.5 has been found for
one of SDSS quasars, SDSSp J104433.04$-$012502.2 at $z=5.74$ (Shioya
et al. 2002).  The detection of such gravitational lensing by a
high-redshift galaxy ($z > 1$) is rare currently.  Since the number
density of potential lensing galaxies could decrease with increasing
redshift, it seems difficult to find events of gravitational lensing
by such high-redshift galaxies.  Relatively lower masses of
high-redshift galaxies also make it difficult to detect them. However,
such events will help us in investigating very high-redshift galaxies
(e.g., $z > 10$) in future very-deep and wide-field imaging surveys.

\acknowledgments

We would like to thank to the staff member of Subaru Telescope.  We
would also thank to anonymous referee for his/her useful comments and
suggestions.  This work was financially supported in part by the Japan
Society for the Promotion of Science (Nos. 17253001, 19340046,
23244031, 23654068 [YT], 23684010 [AKI], and 24244018 [II]).



\end{document}